%% file: servia2015Deciding.tex
\documentclass{sig-alternate-2013}

\usepackage{amsfonts}
\usepackage{amsmath}
\usepackage{amssymb}
\usepackage{algorithm}
\usepackage[noend]{algpseudocode}

\usepackage{url}

\usepackage{booktabs}
\usepackage{tabularx}

\usepackage{color}
\usepackage{soul}
\usepackage{comment}


\usepackage{tikz}
\usetikzlibrary{arrows}

\usepackage[font=bf,labelsep=colon]{caption} 
\DeclareCaptionType{copyrightbox}
\usepackage{graphicx}
\usepackage{subcaption}
\usepackage[caption=false]{subfig}

\begin{document}

\title{Deciding what to display: maximizing the information value of social media}

\numberofauthors{3} 
%
\author{
%
%
\alignauthor
Sandra\titlenote{Corresponding author.}\titlenote{This work was partially done while Sandra Servia-Rodr\'iguez was an intern at HP Labs.} Servia-Rodr\'{i}guez\\
       \affaddr{AtlantTIC, University of Vigo}\\
       \email{sandra@det.uvigo.es}
\alignauthor
Bernardo A. Huberman\\
       \affaddr{HP Labs}\\
       \email{bernardo.huberman@hp.com}
\and  
\alignauthor Sitaram Asur\\
    \affaddr{HP Labs}\\
    \email{sitaram.asur@hp.com}
}

\maketitle
\begin{abstract}
    In information-rich environments, the competition for users' attention leads to a flood of content from which people often find hard to sort out the most relevant and useful pieces. Using Twitter as a case study, we applied an attention economy solution to generate the most informative tweets for its users. By considering the novelty and popularity of tweets as objective measures of their relevance and utility, we used the \emph{Huberman-Wu} algorithm to automatically select the ones that will receive the most attention in the next time interval. Their predicted popularity was confirmed by using Twitter data collected for a period of 2 months.
    
\end{abstract}

\input{Introduction}
\input{Background}
\input{Methodology}

\input{Evaluation}
\input{PeakHours}

\input{Related}

\input{Conclusion}

\section{Acknowledgments}
Work partially funded by the Spanish Ministry of Economy and Competitiveness (EEBB-I-14-07966, TEC2013-47665-C4-3-R); the European Regional Development Fund and the Galician Regional Government under agreement for funding the AtlantTIC Research Center; and the Spanish Government and the European Regional Development Fund under project TACTICA.

\bibliographystyle{abbrv}
\bibliography{biblio}  

\end{document}

%% file: Introduction.tex
\section{Introduction}
\label{sec:introduction}
 
The popularity of the Web and social media services has resulted in a constant flood of information which makes it hard for users to identify and consume the most relevant and useful pieces of content.
Given the limited amount of attention that users can afford, providers of content have to decide what items to prioritize in order to gain the attention of users and become popular.
 
In earlier research, the task of automatically selecting the most relevant and useful pieces of information has been approached from different perspectives. \emph{Ranking}~\cite{baeza1999modern} is at the core of the Information Retrieval (IR) scientific discipline behind search engines as \emph{Google} or \emph{Yahoo! Search}, where pieces of information -documents- are ranked according to their relevance to a given query. \emph{Recommendation}~\cite{resnick1997recommender}, the discipline behind the success of many online services as shopping, photo(music, video)-sharing or online social networks, aims to predict the pieces of information that users will find more useful; either because these pieces are (i) similar to previous pieces liked by them -\emph{content-based filtering}-, or (ii) liked by other users with similar preferences -\emph{collaborative filtering}. However, neither ranking nor recommendation are suitable for deciding the content to prioritize in social media, since the former requires a query to answer and the latter the preferences of the subjects (readers) to receive the recommendation. So far, in online newspapers, magazines and blogs, editors have been the ones to decide the choice of content and the presentation order.
However, the emergence of news media aggregators, such as \emph{\url{digg.com}} or \emph{\url{reddit.com}}, has led to a citizen journalism-based ordering. That is, instead of having professional editors to determine the important news, people vote for news that they find interesting and the votes received by an article play an important role in its ranking with respect to other news on the front page or in the different ordered lists of news.
 
Social media services such as Twitter feature a large number of subscribers and they serve as aggregators of content such as news, promotional campaigns, media and status updates from users. Given the diversity and magnitude of content that is available, it is important, from the service provider's point of view to ensure easy access to relevant information to users, in order to retain and increase user engagement with the platform.
 
The Twitter timeline displays all tweets from the users that subscribers follow in decreasing order of publication. However, novelty is not the only feature that makes tweets valuable to users. Other features such as popularity, can also contribute to give value to tweets. Although Twitter has recently included a feature to display in bigger size those tweets that have received more engagement (retweets, favourites,...), users still have to scroll down their timeline in order to find interesting tweets. We are convinced that, apart from novelty, popularity should play a key role in deciding which tweets should be displayed in the top positions in order to increase users' engagement with the platform. This would specially be useful for those users who log onto Twitter using mobile devices (mobile phones, tablets,...), due to the reduced dimension of their screens (when compared with computers).
 
Within this setting, \emph{we study a method for selecting the optimal arrangement of tweets that maximizes the information value of users}. Considering the number of retweets as an indicator of the popularity of a tweet and the time since it was posted as an indicator of its novelty, we conduct a study to empirically validate the solution proposed by Huberman and Wu in~\cite{huberman2008economics} to obtain the optimal arrangement of tweets that maximizes the informativeness for the users.
By mapping the problem to that of optimal allocation of effort for a number of competing projects, Huberman and Wu formulate the problem as a special case of \emph{bandit problem}~\cite{gittins1979bandit,whittle1988restless}, which they finally solve by applying the adaptive greedy algorithm proposed by Bertsimas and Ni{\~n}o-Mora~\cite{bertsimas1996conservation}.
 
To evaluate our arrangement method, we crawled the Twitter streams of $5$ different influential news media accounts (\emph{New York Times}, \emph{BBC}, \emph{CNN}, \emph{Huffington Post} and \emph{Mashable}) and collected their tweets and retweets for a period of two months. We obtained an arrangement taking into account the popularity and novelty of content and validated this using the actual attention provided by users in the form of retweets, replies and favorites.
Our results show high accuracy of prediction of user attention thus demonstrating the benefits of our proposed solution.
We believe that these findings will be very useful for content providers to automatically organize the most informative items for their customers and, in this way, win their attention.

%% file: Background.tex
\section{Background}
\label{sec:background}

As mentioned, our goal with this research is to demonstrate the suitability of using the approach in \cite{huberman2008economics} to the problem of selecting the most informative tweets to users. This solution involves the steps of (i) mapping the problem of optimizing the information one gets to that of the optimal allocation of effort to a number of competing problems, (ii) formulating the problems as a special type of \emph{bandit problem}, \emph{dual-speed restless bandit problem}, and (iii), by using the adaptive greedy algorithm developed by Bertsimas and Ni{\~n}o-Mora~\cite{bertsimas1996conservation}, calculate an index for each items state, which is then used to decide which item goes into the top list of a given time. Below we detail these steps.

\subsection{Problem formulation}

Consider a system that wishes to present $n$ items to a large group of users but it can only present $k$ ($k < n$) at any given time. Since an item displayed in front of a user has a higher probability of being chosen than when it is not displayed, we will call these $k$ items the ``top list''. We will also assume that the system can update its top list at discrete times $t = 0,1,2,...$.

If the system can track a certain set of properties for each item, such as its reputation, history or age, we say that the item is in a ``state'' defined by those properties. Let $E$ be the set of all possible states, i.e. all possible combination of those trackable properties. In general, the state of an item may change as time goes on. We assume that the state of each item changes according to a Markov process independent of the state of other items, with transition probabilities $\{P_{ij}^{1} : i,j \in E \}$ if the item is on the top list, and $\{P_{ij}^{0} : i,j \in E \}$ if it is not. We also make the assumption that having an item on the top list encourages more users to try it out and consequently accelerates its transitions from one state to the other. Conversely, when an item transitions away from the top list it slows down its rate of change by an amount $\epsilon_{i}$ which is less than $1$. This \emph{dual-speed assumption} can be stated as

\begin{equation}
    P_{ij}^{0} = \left\{
    \begin{array}{cl}
    \epsilon_{i} P_{ij}^{1}, & \ i \neq j \\
    (1-\epsilon_{i}) + \epsilon_{i} P_{ii}^{1}, & \ i = j
    \end{array}
    \right.
\end{equation}

\noindent where $\epsilon_{i} \in [0,1]$.

Consider the total expected utility $r_i$ obtained at one time step by those users who decide to try an item on the top list which has state $i$. This utility may depend on many factors, such as the total expected number of users choosing the item at a given time step, or the expected quality of the item. Since we can always enlarge the definition of ``state'' to include these factors, the utility $r_i$ is uniquely determined by the item state $i$. In other words, we can assume that $r = (r_{i})_{i \in E}$ is an $|E|$-dimensional constant vector known by the system. 

Our goal then is to design a system that maximizes the total expected utility of all users:

\begin{equation}
    \max_{u \in U} \mathbb{E}_u \Bigg [ \displaystyle\sum_{t=0}^{\infty} \displaystyle\sum_{m=1}^{n} \beta^t r_{i_m(t)} I_m(t) \Bigg ] ,
\end{equation} 

\noindent where $0 < \beta \leq 1$ is the future discount factor, $i_m(t)$ is the state of item $m$ at time $t$, and

\begin{equation}
    I_m(t) = \left\{
    \begin{array}{cl}
    1 & \ \text{if item } m \text{ is displayed at time }t , \\
    0 & \ \text{otherwise.}
    \end{array}
    \right.
\end{equation}

\noindent We seek to find the optimal strategy, $u$, in the space $U$ of stationary strategies (strategies that depend on current item states only). This optimal strategy can then get translated into the set of offerings that should appear in the top list.

\subsection{Solution}

The model described is a \emph{dual-speed restless bandit} problem: \emph{restless} because changes of state can also occur when the items are not displayed in the top list and \emph{dual-speed} because those changes do happen at a different speed than those on the top list. Bertsimas and Ni{\~n}o-Mora showed that a \emph{relaxed} version of the dual-speed problem is always indexable --it is possible to attach an \emph{index} to each item state, so that the top list is the one including those items with the largest indices-- and proposed an efficient adaptive greedy algorithm to compute these indices~\cite{bertsimas1996conservation}.

Before using the algorithm, it is necessary to calculate a set of constants $A_i^S$. Assuming that $E$ is finite, for any subset $S \subseteq E$, we define the $S$-active policy $u_S$ to be the strategy that recommends all items whose state is in $S$. Considering an item that starts from an initial state $X(0) =i$, under the action implied by strategy $u_S$, its total occupancy time in $S$ is given by

\begin{equation}
    V_i^S = \mathbb{E}_{u_S} \Bigg [ \displaystyle\sum_{t=0}^{\infty} \beta^t I_S(t)|X(0) = i \Bigg ] ,
\end{equation}

\noindent where

\begin{equation}
    I_S(t) = \left\{
    \begin{array}{cl}
    1 & \ \text{if } X(t) \in S, \\
    0 & \ \text{otherwise.}
    \end{array}
    \right.
\end{equation}

\noindent We have

\begin{equation}
    V_i^S = \left\{
    \begin{array}{cl}
    1+ \beta \displaystyle\sum_{j \in E} P_{ij}^1 V_j^S ,  & \ i \in S, \\
    \beta \displaystyle\sum_{j \in E} P_{ij}^0 V_j^S ,  & \ i \in S^c. 
    \end{array}
    \right.
\end{equation}

\noindent The variables $\{ V_i^S \}_{i \in E}$ can be solved from the set of linear equations above. 

A matrix of constants $\{ A_i^S \}_{i \in E, S \subseteq E}$ is defined by means of $V_i^S$, as follows:

\begin{equation}
    A_i^S = 1 + \beta \displaystyle\sum_{j \in E} P_{ij}^1 V_j^{S^c} - \beta \displaystyle\sum_{j \in E} P_{ij}^0 V_j^{S^c}.  
\end{equation}

\noindent The constants $\{ A_i^S \}$ are then used in the Bertsimas-Ni{\~n}o-Mora algorithm as indicated in Algorithm~\ref{bertsimasNinoMora}.

\footnotesize\begin{algorithm}
\caption{Bertsimas-Ni{\~n}o-Mora adaptive greedy algorithm}\label{bertsimasNinoMora}
\begin{algorithmic}[1]
\Statex \textbf{Step 1.} Set $S_{|E|} = E$ and
\begin{equation}
 y^{S_{|E|}} =  \max \bigg \{ \frac{r_i}{A_i^E}: i \in E \bigg \}.
\end{equation}
\Statex Select $\pi_{|E|}$ as any maximizer and set $G_{\pi_{|E|}} = y^{S_|E|}$.
\Statex 
\Statex \textbf{Step 2.} For $k = 2,3,...,|E|$, set $S_{|E|-k+2} \setminus \{ \pi_{|E|-k+2} \}$ and 
\begin{align}
 y^{S_{|E|-k+1}} =  \max &  \Bigg   \{  \frac{r_i - \sum_{j=1}^{k-1} A_i^{S_{|E|-j+1}} y^{|E|-k+1} }{A_i^{S_{|E|-k+1}}} \nonumber \\
                                & :  i \in S_{|E|-k+1} \Bigg \}.
\end{align}
\Statex Select $\pi_{|E|-k+1}$ as any maximizer and set $G_{\pi_{|E|-k+1}} = G_{\pi_{|E|-k+2}} + y^{S_|E|-k+1}$.
\end{algorithmic}
\end{algorithm}
\normalsize

Finally, the strategy is to always display the $k$ items whose states have the largest $G$ index.

%% file: Methodology.tex
\section{Deciding what to display on Twitter} 
\label{sec:methodology}

On Twitter, the home timeline is a long stream showing all tweets from those that users have chosen to follow, displayed, at a given time, in decreasing order of publication. Focusing on a particular group of users, influential news media, we are interested on \emph{selecting the optimal arrangement of tweets to be displayed to their followers} in order to maximize their informational value and, in this way, grab users' attention.

With this aim, we particularize the \emph{Huberman-Wu} algorithm (Section~\ref{sec:background}) to this scenario. In order to properly define the states and compute the transition probabilities between these states, we use actual data from Twitter. Specifically, we have been monitoring the Twitter accounts of $5$ different news media and the retweets done to their tweets for a period of $2$ months. In what follows, we describe the dataset obtained and how we use the tweets of the first month to set up the states and transition probabilities and the data of the last month to prove the suitability of our approach (Section~\ref{sec:evaluation}).

\subsection{Dataset}
\label{sec:meth-data}

\begin{figure*}
	\begin{tabular}{cc}
    \begin{subfigure}[b]{0.5\textwidth} 
        \centering
        \includegraphics[scale=.45, angle=0]{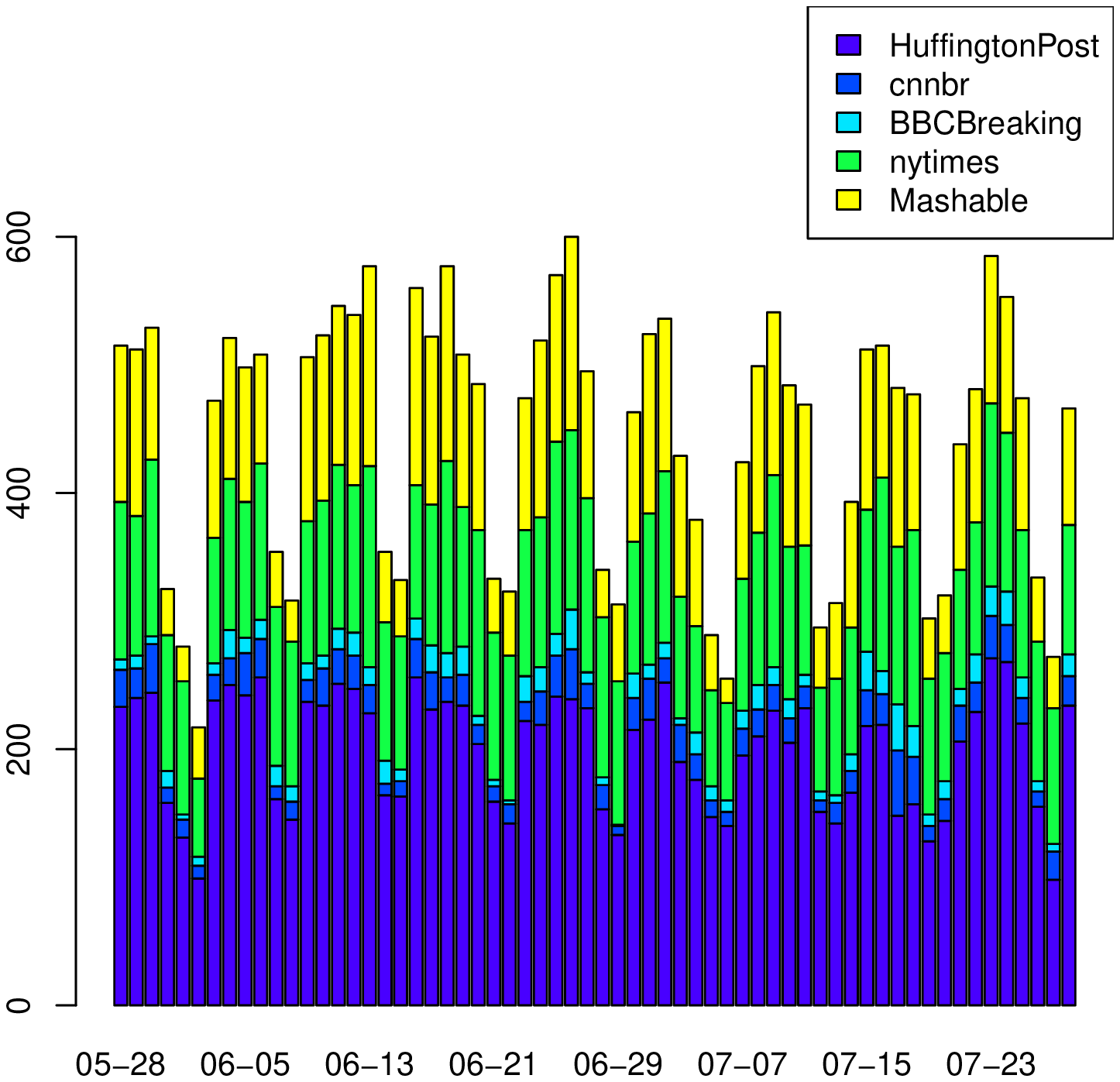}
        \caption{\# of tweets posted per day during the $2$ months of observation.}
        \label{fig:tweetsTime_a}
    \end{subfigure} &
    \begin{subfigure}[b]{0.5\textwidth}
        \centering
        \includegraphics[scale=.45, angle=0]{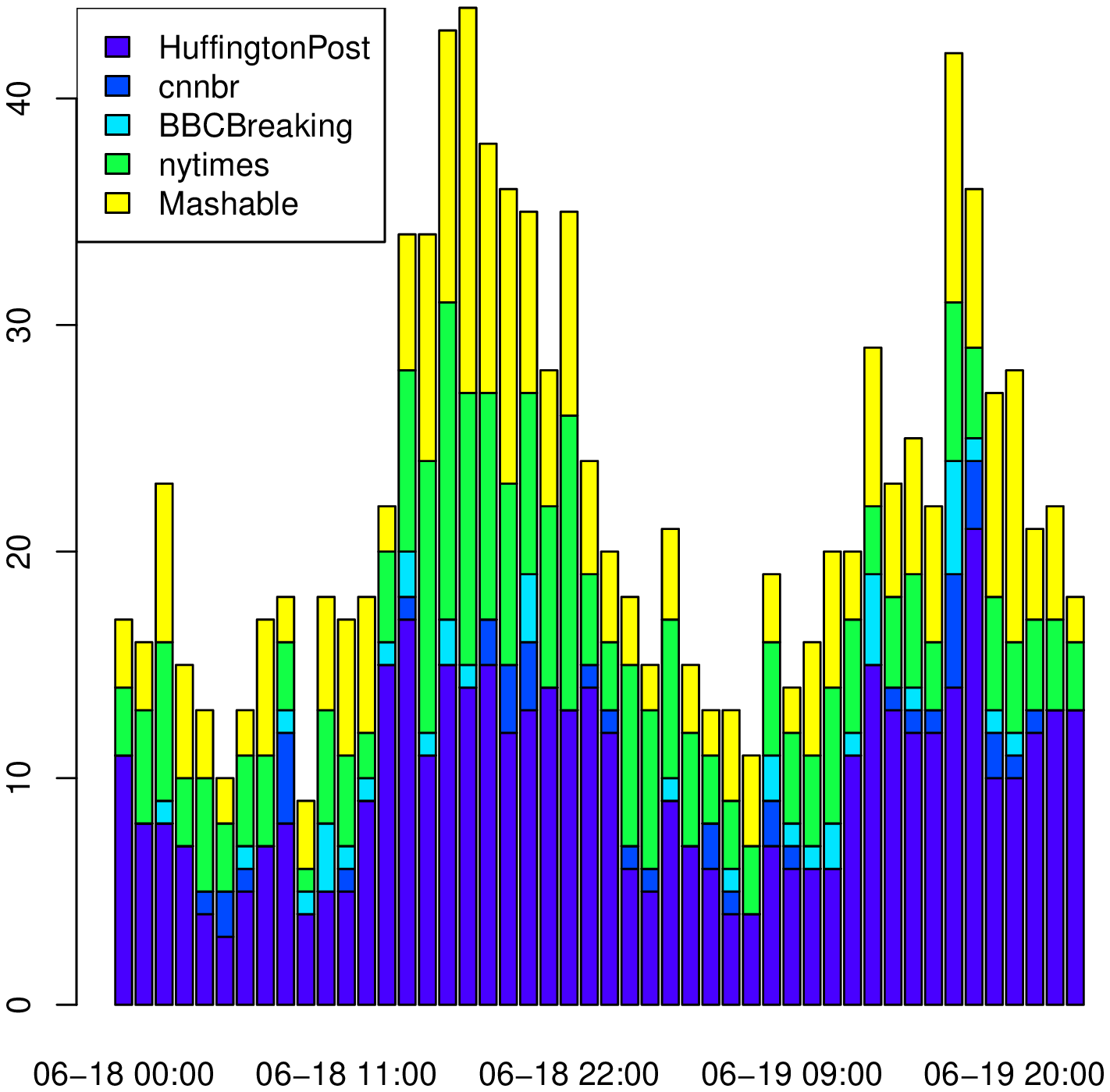}
        \caption{\# of tweets posted per hour during $2$ days of observation.}
        \label{fig:tweetsTime_b}
    \end{subfigure} \\ 
    \end{tabular}
    \caption{Temporal distribution of tweets publication.}
      \label{fig:tweetsTime}
\end{figure*}

Our dataset consists of tweets and retweets collected from crawling the Twitter streams of $5$ different influential news media accounts (\emph{The New York Times}, \emph{BBC Breaking News}, \emph{CNN Breaking News}, \emph{Huffington Post} and \emph{Mashable}) for a period of two months. We ended up with a total of $27.548$ tweets and $2.576.853$ retweets to these tweets. Figure~\ref{fig:tweetsTime} shows the temporal distribution of the tweets posted by each news media account. Specifically, Figure~\ref{fig:tweetsTime_a} shows the number of tweets posted per day during the whole period, whereas Figure~\ref{fig:tweetsTime_b} shows the number of tweets posted per hour during one week of observation. Focusing in Figure~\ref{fig:tweetsTime_a}, we observe certain periodicity in the number of tweets posted by our news media per day, seeing that the number of tweets on weekends is lower than on weekdays, and the day with less tweets is Sunday. However, the day when the most tweets are posted varies from one week to the other. In the case of number of tweets per hour (Figure~\ref{fig:tweetsTime_b}), we also observe a periodic behaviour, where most of the tweets are posted between $12am$ and $2am$. Given that the scale is in $UTC+0$, the hours in which less tweets are posted correspond to night hours in the US. Finally, focusing on individual news media sources, there is clearly a gap between the number of tweets posted by \emph{Huffington Post}, \emph{Mashable} and \emph{The New York Times} with respect to \emph{BBC} and \emph{CNN Breaking News}. This is because we consider the accounts of \emph{BBC} and \emph{CNN} that only post \emph{breaking news} (news very popular that receive high engagement from their followers), whereas for the others, we consider their regular accounts.

\subsubsection{Time-dependence of retweets}

\begin{figure*}[t!]
	\begin{tabular}{ccc}
    \begin{subfigure}[b]{0.33\textwidth}
        \centering
        \includegraphics[scale=.32, angle=0]{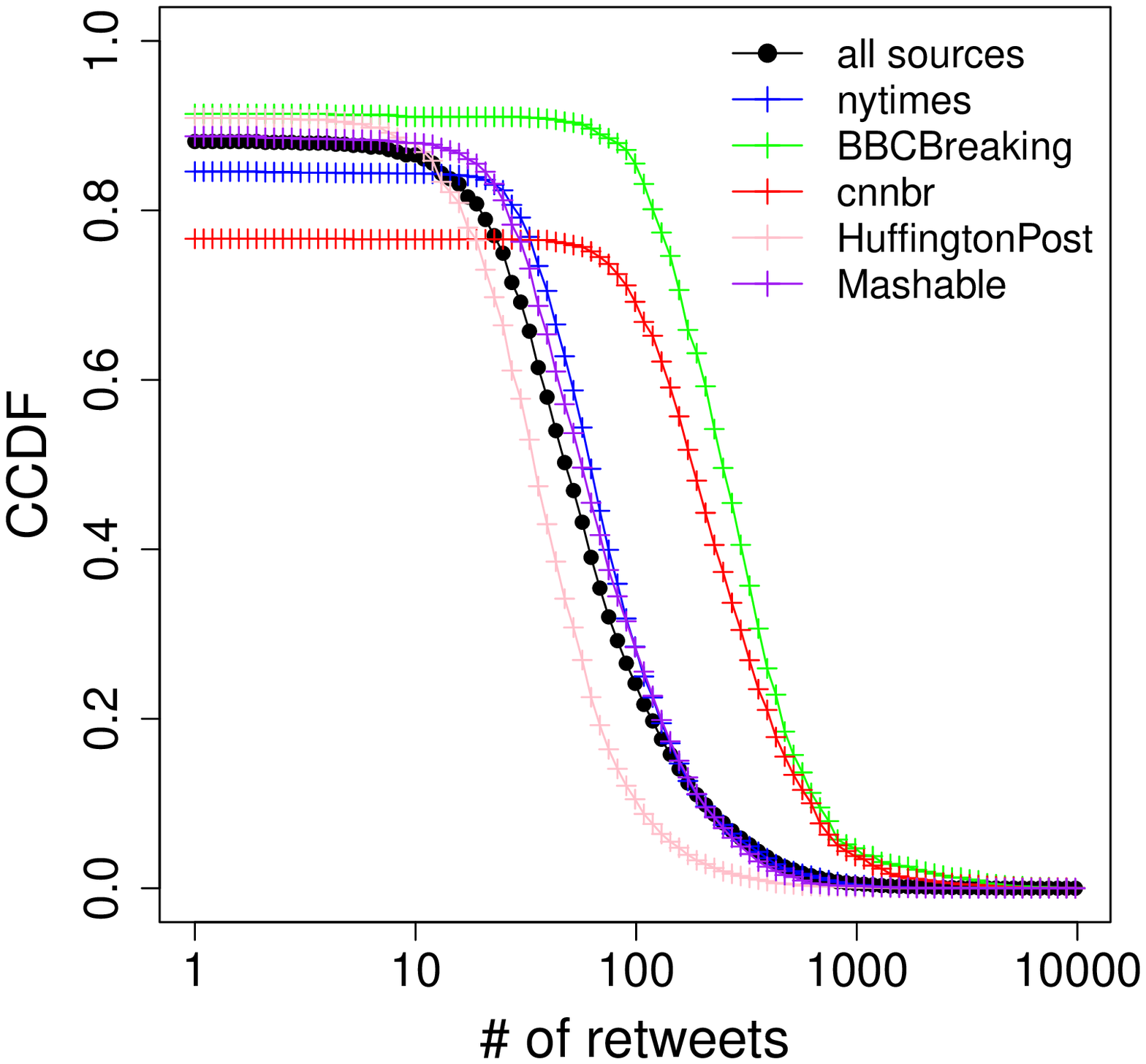}
        \caption{CCDF of the \# of retweets received per tweet for each one of the news accounts considered.}
        \label{fig:tweetsRT_a}
    \end{subfigure} &
    \begin{subfigure}[b]{0.33\textwidth}
        \centering
        \includegraphics[scale=.32, angle=0]{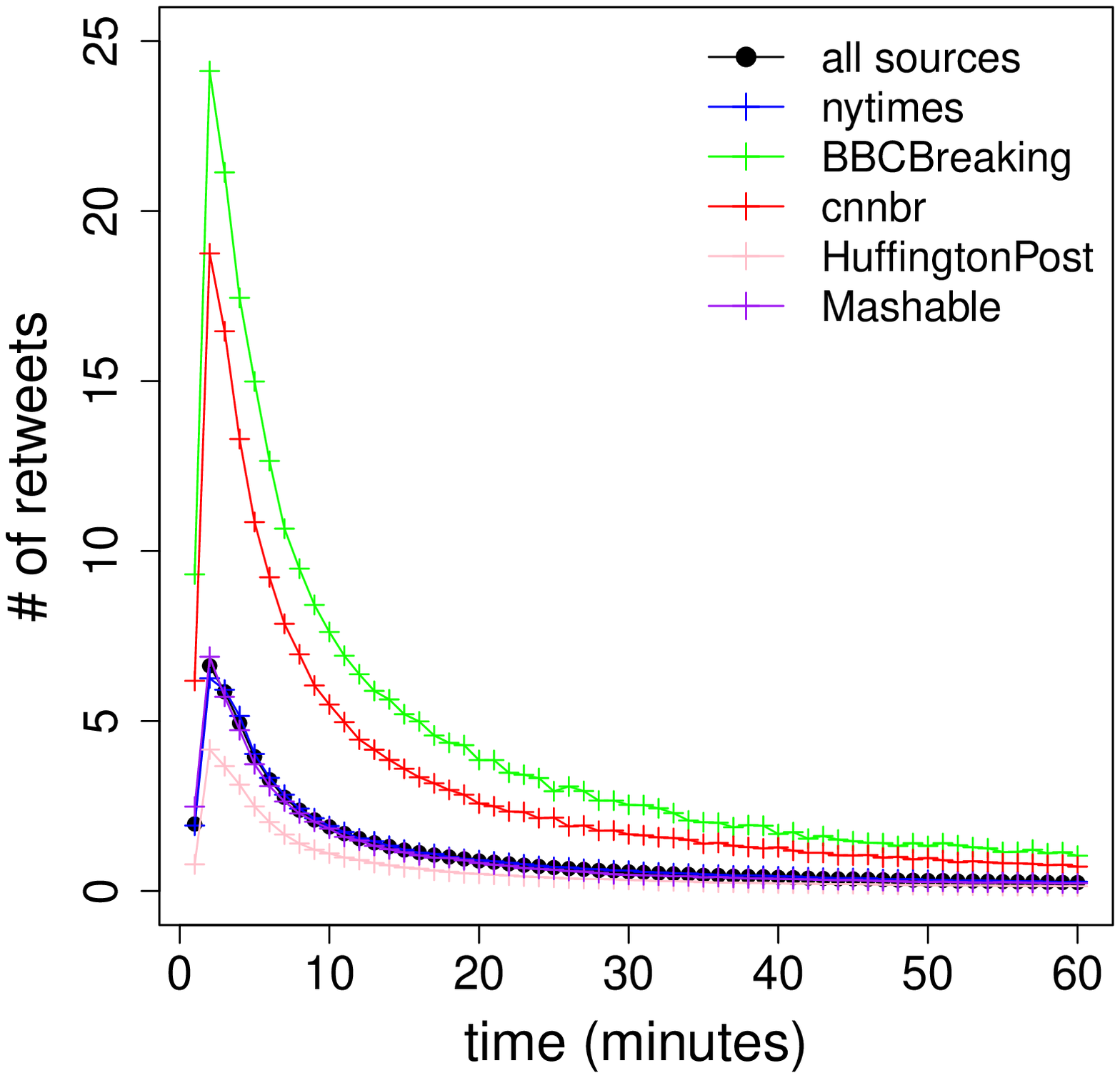}
        \caption{Average \# of retweets received per tweet after \emph{time} minutes of its publication (split by news account)}
        \label{fig:tweetsRT_b}
    \end{subfigure} &
    \begin{subfigure}[b]{0.33\textwidth}
        \centering
        \includegraphics[scale=.32, angle=0]{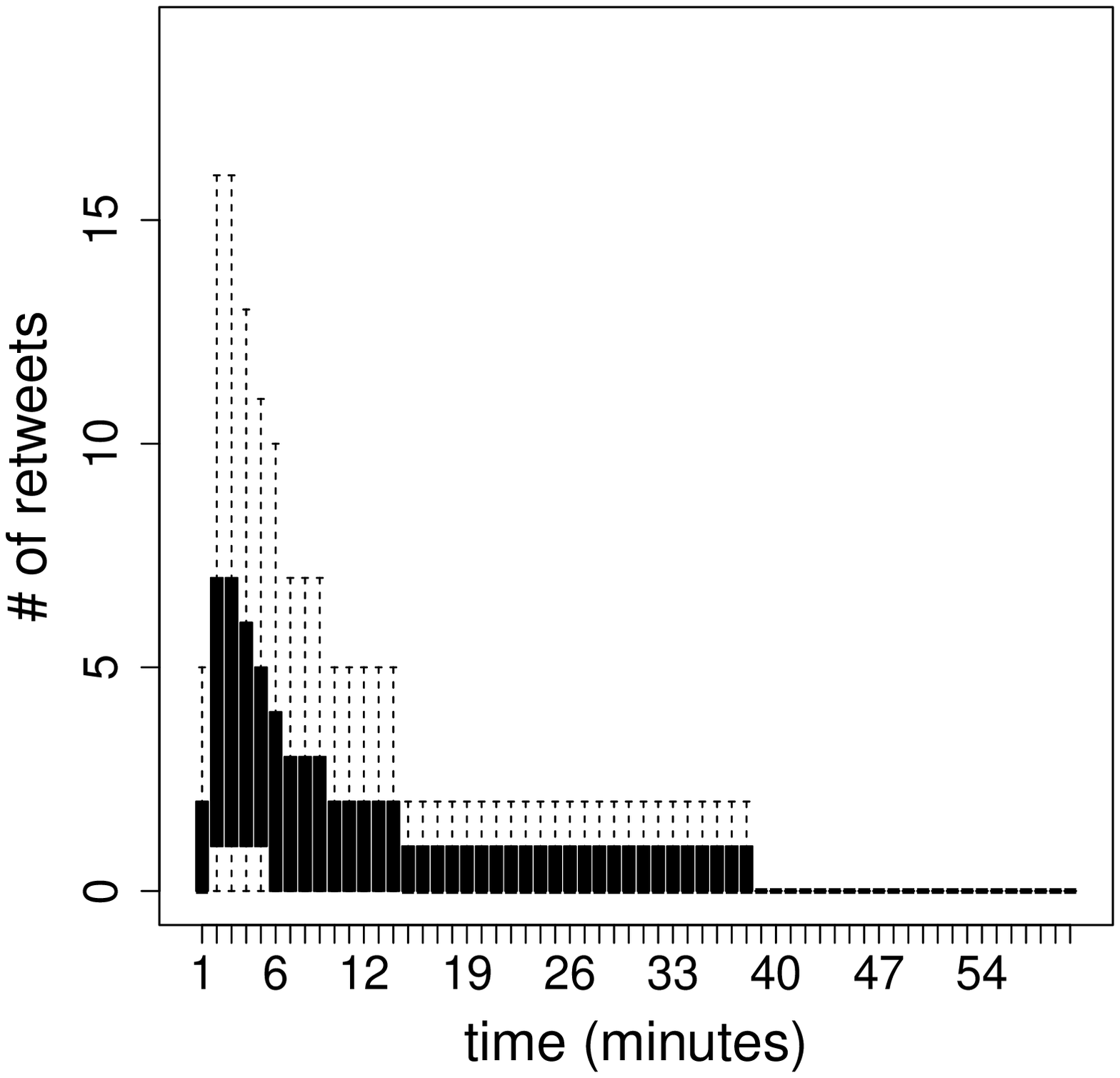}
        \caption{Scattering of the \# of retweets received per tweet after \emph{time} minutes of its publication for all the tweets in the dataset.}
        \label{fig:tweetsRT_c}
    \end{subfigure} \\
    \end{tabular}
    \caption{(Temporal) distribution of retweets per tweet.}
      \label{fig:tweetsRT}
\end{figure*}

So far, we have described the dataset in terms of tweets posted by news media accounts, but now we focus on the retweets to these tweets. And specifically, in the number of retweets that every tweet has received (Figure~\ref{fig:tweetsRT_a}) and the time between the publication of tweets and their retweets (Figure~\ref{fig:tweetsRT_b}). Although the majority of tweets posted by our news media receive some retweets, there are huge differences in the number of retweets received by them. For instance, \emph{BBC} and \emph{CNN} are the ones that receive more engagement from their users, mainly because they only post breaking news (popular news). On the contrary, \emph{Huffington Post}, apart from being the news media that publishes the most tweets, is the one whose tweets receive the least engagement (less retweets) from their followers. 

Figure~\ref{fig:tweetsRT_b} contains the temporal distribution of the average number of retweets per tweet in the hour after their publication. Here we see that, independent of the news media considered, (i) tweets get more engagement in the second and third minute after their publication and (ii) since the second minute, the number of retweets achieved fits a power law distribution. Therefore, what is clear is that the most recent tweets, i.e. those in the top-positions of the timeline, are the most exposed to the users and have more possibilities of being retweeted. Finally, Figure~\ref{fig:tweetsRT_c} shows that the scattering, and therefore the variance, of the number of retweets received between $2$ and $9$ minutes after publication is higher than after $10$ minutes, with the number decreasing significantly after $38$ minutes from the tweet publication. This shows that, with the exception of the retweets produced in the first minute after publication, the variance of the number of retweets received per minute decreases over time.

\subsubsection{Temporal comparison with other platforms} 

In our system, we only consider tweets posted by news media and therefore most of them contain news with limited lifetimes. For this reason, the time difference between a retweet and its original tweet is very less, and the majority of retweets occur within the first hour of publication of the original tweet. However, when considering the whole of Twitter, only half of the retweets are made during this first hour, whereas the rest can even happen a month after the publication of the original tweet~\cite{kwak2010twitter}. Although we selected Twitter for our study, other systems providers of content present a similar tendency in terms of the drop of the attention received with the loss of novelty. This is the case of the interactive website \emph{\url{digg.com}}, in which the attention received by the stories that users discover from the Internet and upload to the platform decay with their loss of novelty~\cite{wu2007novelty}. However, this reduction, at least for popular stories, is not as noticeable as in the case of Twitter, since this platform (\emph{digg}) prioritizes the information to display taking into account their novelty, but also their popularity. So, the more times popular items are displayed, the more prone they are to get users' attention. As a conclusion, in Twitter, and especially when considering news, the decay is faster than in other platforms, and therefore the interval between discrete times in which our system can update the display should be chosen to be short.

\subsubsection{Conditional variance}
We also calculated the conditional variance of the number of retweets received after $t$ minutes from the publication of their original tweets. This is the variance in terms of number of retweets received after $t$ minutes from publication by those tweets that had received the same number of retweets after $t-1$ minutes. Our results revealed variance values that were larger than zero for all the different values of retweet counts in $t$.

\subsection{Setting up the algorithm parameters}
\label{sec:meth-something}

\subsubsection{States and reward}
\label{sec:st-rew}

In Twitter, we can track a certain set of properties for each tweet, such as age, number of retweets, favourites, etc. We consider that the properties that define the state of each item -tweet- at each instant $t$ are its \emph{novelty} -time since publication- and \emph{popularity} -understanding this as the \emph{number of retweets} that it has received-. In order to have a finite set of states $E$, we discretize the possible values of novelty and number of retweets, ending up with $10$ different values for \emph{novelty} and $10$ for \emph{popularity}. Hence each state can be represented as a $2$-vector $(n,p) \in \{ 1,2,3,4,5,6,7,8,9,10 \}$, where $n$ is the novelty -time since publication- and $p$ is the popularity -number of retweets received-. In addition to those $100$ states, we also consider the state $0$, the ``unknown state''. Each item initially starts in this state and also ends on it, with this state serving as both the sink and the source.

\begin{figure*}[t!]
	\begin{tabular}{cc}
    \begin{subfigure}[b]{0.5\textwidth} 
        \centering
        \includegraphics[scale=.45, angle=0]{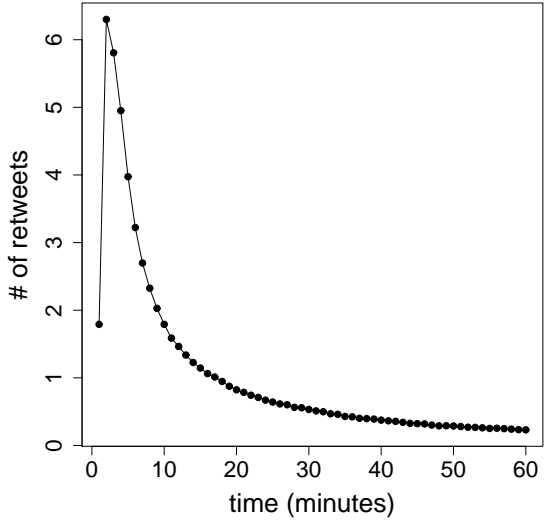}
        \caption{Average \# of retweets received per tweet after \emph{time} minutes of its publication for all the tweets posted during the first month of observation.}
        \label{fig:tweetsRTtraining_a}
    \end{subfigure} &
    \begin{subfigure}[b]{0.5\textwidth}
        \centering
        \includegraphics[scale=.45, angle=0]{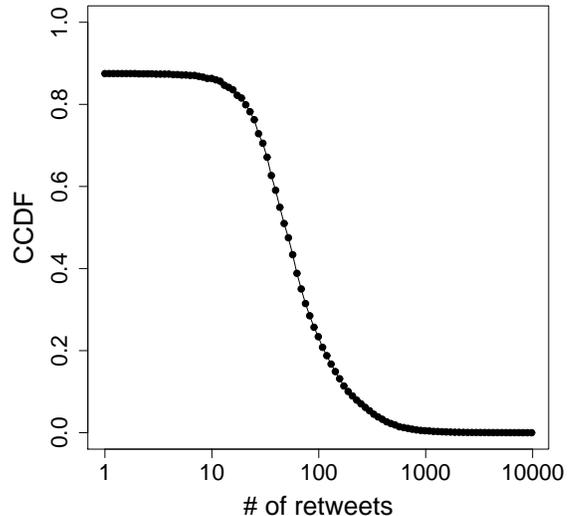}
        \caption{CCDF of the \# of retweets received per tweet for all the tweets posted during the first month of observation.}
        \label{fig:tweetsRTtraining_b}
    \end{subfigure} \\ 
    \end{tabular}
    \caption{(Temporal) distribution of retweets per tweet during the first month of observation.}
      \label{fig:tweetsRTtraining}
\end{figure*}

In order to set the reward and the values of the properties that define each state we look at the \emph{novelty} and \emph{popularity} of the tweets posted during the first month of observation. Figure~\ref{fig:tweetsRTtraining_a} shows the average number of retweets per minute that tweets receive during the hour immediately after their publication, whereas Figure~\ref{fig:tweetsRTtraining_b} represents the CCDF (Complementary Cumulative Distribution Function) of the number of retweets received by those tweets posted during the \emph{first month} of observation. Focusing on Figure~\ref{fig:tweetsRTtraining_a}, we observe that, on average, tweets receive the majority of retweets in the first minutes after publication and it is also in these first minutes when there are the highest differences between the average number of retweets received in one interval with respect to the others. Therefore, we decided to set the limits between the different intervals that define the state to

\begin{equation}
    lim_n = \{ 1,2,3,4,5,6,7,8,9,20,60 \}.
\end{equation}

\noindent So, the state of novelty $i \in n$ contains the tweets that were posted between $lim_n[i]$ and $lim_n[i+1]-1$ minutes before the current time of observation. 

Focusing on the popularity of the tweets --\emph{number of retweets received}-- (Figure~\ref{fig:tweetsRTtraining_b}), we observe that the the number of retweets per tweet is distributed according to a power law distribution, where the majority of the tweets receive less than $100$ retweets whereas a very small percentage of tweets is retweeted more than $1000$ times. In order to set the popularity of the states, we split the tweets, sorted according to the times they are retweeted, into equal sized subsets. So, the limits between the different intervals that define the state are

\begin{equation}
    lim_p = \{ 0,1,19,25,32,39,48,61,82,131, \infty \}.
\end{equation}

\noindent So, the state of popularity $j \in p$ contains the tweets that have been retweeted between $lim_p[j]$ and $lim_p[j+1]-1$ times before the current time of observation.

Finally, we reward the tweets in proportion to the number of retweets they receive and their novelty. So, we set the reward of each state to

\begin{equation}
    \left\{
       \begin{array}{ccl}
       r(n,p) & = & r_{n} * r_{p} ,  \\
       r(0) & = & 0
       \end{array}
    \right.
\end{equation}

\noindent where the $r_n$ and $r_p$ are the normalized average number of retweets per interval. That is, the average number of retweets received between $lim_n[i]$ and $lim_n[i+1]-1$ minutes after publication in the case of novelty, and the average number of total retweets received by those tweets that have received between $lim_p[i]$ and $lim_p[i+1]-1$ retweets in the case of popularity. This results in

\begin{equation}
\begin{array}{ccl}
r_{n} & = & \{0.28, 1, 0.92, 0.79, 0.63, 0.51, 0.43, 0.37, 0.21, 0.07 \} , \\
r_{p} & = & \{0.01, 0.07, 0.11, 0.16, 0.19, 0.24, 0.28, 0.34, 0.44, 1 \} .
\end{array}
\end{equation}

\noindent Please, note that the reward when $p=1$ is not zero but, in order to conserve the reward of the novelty in $r(n,1) / n \in \{ 1,2,3,4,5,6,7,8,9,10 \}$, we consider that the average number of retweets in this set is $1$.

\subsubsection{Transition probabilities}

As mentioned, we assume that the state of each item changes according to a Markov process independent of the state of other items, with transition probabilities $\{ P_{ij}^1 : i,j \in E \}$ if the item is on the top list, and $\{ P_{ij}^0 : i,j \in E \}$ if it is not. In order to empirically calculate these transition probabilities we consider all the tweets posted during the first month of observation. Assuming that all the items are on the top list (all of them are displayed), we define $\{ P_{ij}^1 : i,j \in E \}$ as

\begin{equation}
    P_{ij}^1 = \frac{|I_{j}(t+1)|}{|I_{i}(t)|} 
\end{equation}

\noindent where $I_i(t)$ is the set of items in state $i$ at time $t$ and $I_j(t+1)$ the set of items in state $j$ at $t+1$ that transited to this state from state $i$. Finally, we fix $\epsilon_i = 0.1$ for all $i \in E$, which expresses the fact that displaying an item on the top list accelerates its transition speed by ten times.

\subsection{Solution}
\label{sec:meth-solution}

The $G$ index rankings of the $101$ states are calculated using the Bertsimas-Ni{\~n}o-Mora heuristic described in Section~\ref{sec:methodology}. The results are shown in Figure~\ref{fig:graph}. As can be seen from the figure, state $(2,10)$ has the largest $G$ index, state $(10,3)$ the second-largest, and so on. As mentioned, the absolute value of the indices are not as important as their relative orders: items --tweets-- should be displayed according to the relative order of the indices of their states.

The result is by no means trivial. For example, the top state (1) is not the most novel but it is the most popular. On the other hand (6) is less popular but more novel than (7). Also, the fact that the algorithm gives high index values to potentially valuable states means that the unknown state which gives no reward should have higher display priority than other states with positive reward. Finally, note also that the influence of the popularity in the output is higher than the novelty, which supports our premise that the current novelty itself doesn't maximise the expected value (informativeness) for the user.

\input{graph}

%% file: graph.tex
\vspace{1cm}

\begin{figure*}
\centering
\begin{tikzpicture}[scale=1.5,every node/.style={draw,circle}]
    
    \node (1) at (0,1) {$89$};
    \node (2) at (0,2) {$92$};
    \node (3) at (0,3) {$101$};
    \node (4) at (0,4) {$100$};
    \node (5) at (0,5) {$99$};
    \node (6) at (0,6) {$98$};
    \node (7) at (0,7) {$97$};
    \node (8) at (0,8) {$96$};
    \node (9) at (0,9) {$95$};
    \node (10) at (0,10) {$94$};
    
    \node (11) at (1,1) {$86$};
    \node (12) at (1,2) {$87$};
    \node (13) at (1,3) {$90$};
    \node (14) at (1,4) {$81$};
    \node (15) at (1,5) {$77$};
    \node (16) at (1,6) {$72$};
    \node (17) at (1,7) {$68$};
    \node (18) at (1,8) {$63$};
    \node (19) at (1,9) {$60$};
    \node (20) at (1,10) {$91$};
    
    \node (21) at (2,1) {$88$};
    \node (22) at (2,2) {$83$};
    \node (23) at (2,3) {$74$};
    \node (24) at (2,4) {$70$};
    \node (25) at (2,5) {$66$};
    \node (26) at (2,6) {$59$};
    \node (27) at (2,7) {$52$};
    \node (28) at (2,8) {$45$};
    \node (29) at (2,9) {$43$};
    \node (30) at (2,10) {$80$};
    
    \node (31) at (3,1) {$84$};
    \node (32) at (3,2) {$76$};
    \node (33) at (3,3) {$65$};
    \node (34) at (3,4) {$62$};
    \node (35) at (3,5) {$55$};
    \node (36) at (3,6) {$48$};
    \node (37) at (3,7) {$40$};
    \node (38) at (3,8) {$36$};
    \node (39) at (3,9) {$31$};
    \node (40) at (3,10) {$71$};
    
    \node (41) at (4,1) {$85$};
    \node (42) at (4,2) {$73$};
    \node (43) at (4,3) {$56$};
    \node (44) at (4,4) {$53$};
    \node (45) at (4,5) {$47$};
    \node (46) at (4,6) {$39$};
    \node (47) at (4,7) {$32$};
    \node (48) at (4,8) {$27$};
    \node (49) at (4,9) {$24$};
    \node (50) at (4,10) {$67$};
    
    \node (51) at (5,1) {$82$};
    \node (52) at (5,2) {$69$};
    \node (53) at (5,3) {$51$};
    \node (54) at (5,4) {$56$};
    \node (55) at (5,5) {$41$};
    \node (56) at (5,6) {$33$};
    \node (57) at (5,7) {$26$};
    \node (58) at (5,8) {$21$};
    \node (59) at (5,9) {$18$};
    \node (60) at (5,10) {$61$};
    
    \node (61) at (6,1) {$79$};
    \node (62) at (6,2) {$64$};
    \node (63) at (6,3) {$44$};
    \node (64) at (6,4) {$42$};
    \node (65) at (6,5) {$35$};
    \node (66) at (6,6) {$28$};
    \node (67) at (6,7) {$20$};
    \node (68) at (6,8) {$17$};
    \node (69) at (6,9) {$14$};
    \node (70) at (6,10) {$54$};
    
    \node (71) at (7,1) {$78$};
    \node (72) at (7,2) {$57$};
    \node (73) at (7,3) {$37$};
    \node (74) at (7,4) {$34$};
    \node (75) at (7,5) {$29$};
    \node (76) at (7,6) {$22$};
    \node (77) at (7,7) {$16$};
    \node (78) at (7,8) {$12$};
    \node (79) at (7,9) {$11$};
    \node (80) at (7,10) {$49$};
    
    \node (81) at (8,1) {$75$};
    \node (82) at (8,2) {$50$};
    \node (83) at (8,3) {$30$};
    \node (84) at (8,4) {$25$};
    \node (85) at (8,5) {$19$};
    \node (86) at (8,6) {$15$};
    \node (87) at (8,7) {$10$};
    \node (88) at (8,8) {$8$};
    \node (89) at (8,9) {$6$};
    \node (90) at (8,10) {$38$};
    
    \node (91) at (9,1) {$58$};
    \node (92) at (9,2) {$23$};
    \node (93) at (9,3) {$9$};
    \node (94) at (9,4) {$7$};
    \node (95) at (9,5) {$5$};
    \node (96) at (9,6) {$4$};
    \node (97) at (9,7) {$3$};
    \node (98) at (9,8) {$2$};
    \node (99) at (9,9) {$1$};
    \node (100) at (9,10) {$13$};

    \node (101) at (5,-1) {$93$};

     \node[draw=none,fill=none] (a) at (-1,0) {};
     \node[draw=none,fill=none] (b) at (-1,1.5) {};
     \node[draw=none,fill=none] (c) at (0.5,0) {};
    
    \path[<->,font=\scriptsize,>=latex] 
    (1) edge (2)
    (2) edge (3)
    (3) edge (4)
    (4) edge (5)
    (5) edge (6)
    (6) edge (7)
    (7) edge (8)
    (8) edge (9)
    (9) edge (10)
    (11) edge (12)
    (12) edge (13)
    (13) edge (14)
    (14) edge (15)
    (15) edge (16)
    (16) edge (17)
    (17) edge (18)
    (18) edge (19)
    (19) edge (20)
    (21) edge (22)
    (22) edge (23)
    (23) edge (24)
    (24) edge (25)
    (25) edge (26)
    (26) edge (27)
    (27) edge (28)
    (28) edge (29)
    (29) edge (30)
    (31) edge (32)
    (32) edge (33)
    (33) edge (34)
    (34) edge (35)
    (35) edge (36)
    (36) edge (37)
    (37) edge (38)
    (38) edge (39)
    (39) edge (40)
    (41) edge (42)
    (42) edge (43)
    (43) edge (44)
    (44) edge (45)
    (45) edge (46)
    (46) edge (47)
    (47) edge (48)
    (48) edge (49)
    (49) edge (50)
    (51) edge (52)
    (52) edge (53)
    (53) edge (54)
    (54) edge (55)
    (55) edge (56)
    (56) edge (57)
    (57) edge (58)
    (58) edge (59)
    (59) edge (60)
    (61) edge (62)
    (62) edge (63)
    (63) edge (64)
    (64) edge (65)
    (65) edge (66)
    (66) edge (67)
    (67) edge (68)
    (68) edge (69)
    (69) edge (70)
    (71) edge (72)
    (72) edge (73)
    (73) edge (74)
    (74) edge (75)
    (75) edge (76)
    (76) edge (77)
    (77) edge (78)
    (78) edge (79)
    (79) edge (80)
    (81) edge (82)
    (82) edge (83)
    (83) edge (84)
    (84) edge (85)
    (85) edge (86)
    (86) edge (87)
    (87) edge (88)
    (88) edge (89)
    (89) edge (90)
    (91) edge (92)
    (92) edge (93)
    (93) edge (94)
    (94) edge (95)
    (95) edge (96)
    (96) edge (97)
    (97) edge (98)
    (98) edge (99)
    (99) edge (100)
    
    (1) edge (11)
    (11) edge (21)
    (21) edge (31)
    (31) edge (41)
    (41) edge (51)
    (51) edge (61)
    (61) edge (71)
    (71) edge (81)
    (81) edge (91)
    (2) edge (12)
    (12) edge (22)
    (22) edge (32)
    (32) edge (42)
    (42) edge (52)
    (52) edge (62)
    (62) edge (72)
    (72) edge (82)
    (82) edge (92)
    (3) edge (13)
    (13) edge (23)
    (23) edge (33)
    (33) edge (43)
    (43) edge (53)
    (53) edge (63)
    (63) edge (73)
    (73) edge (83)
    (83) edge (93)
    (4) edge (14)
    (14) edge (24)
    (24) edge (34)
    (34) edge (44)
    (44) edge (54)
    (54) edge (64)
    (64) edge (74)
    (74) edge (84)
    (84) edge (94)
    (5) edge (15)
    (15) edge (25)
    (25) edge (35)
    (35) edge (45)
    (45) edge (55)
    (55) edge (65)
    (65) edge (75)
    (75) edge (85)
    (85) edge (95)
    (6) edge (16)
    (16) edge (26)
    (26) edge (36)
    (36) edge (46)
    (46) edge (56)
    (56) edge (66)
    (66) edge (76)
    (76) edge (86)
    (86) edge (96)
    (7) edge (17)
    (17) edge (27)
    (27) edge (37)
    (37) edge (47)
    (47) edge (57)
    (57) edge (67)
    (67) edge (77)
    (77) edge (87)
    (87) edge (97)
    (8) edge (18)
    (18) edge (28)
    (28) edge (38)
    (38) edge (48)
    (48) edge (58)
    (58) edge (68)
    (68) edge (78)
    (78) edge (88)
    (88) edge (98)
    (9) edge (19)
    (19) edge (29)
    (29) edge (39)
    (39) edge (49)
    (49) edge (59)
    (59) edge (69)
    (69) edge (79)
    (79) edge (89)
    (89) edge (99)
    (10) edge (20)
    (20) edge (30)
    (30) edge (40)
    (40) edge (50)
    (50) edge (60)
    (60) edge (70)
    (70) edge (80)
    (80) edge (90)
    (90) edge (100);
    
    \path[->,font=\scriptsize,>=latex] 
    (1) edge (101)
    (11) edge (101)
    (21) edge (101)
    (31) edge (101)
    (41) edge (101)
    (51) edge (101)
    (61) edge (101)
    (71) edge (101)
    (81) edge (101)
    (91) edge (101)

    (a) edge node[draw=none,fill=none,auto=right] {$novelty$} (b)
    (a) edge node[draw=none,fill=none,auto=below,yshift=-2ex] {$popularity$} (c);
    
\end{tikzpicture}
\caption{The 101 states ranked by their $G$ indices, ranked from highest to lowest, plus the state $0$ (ranked in position $93$).} 
\label{fig:graph}
\end{figure*}
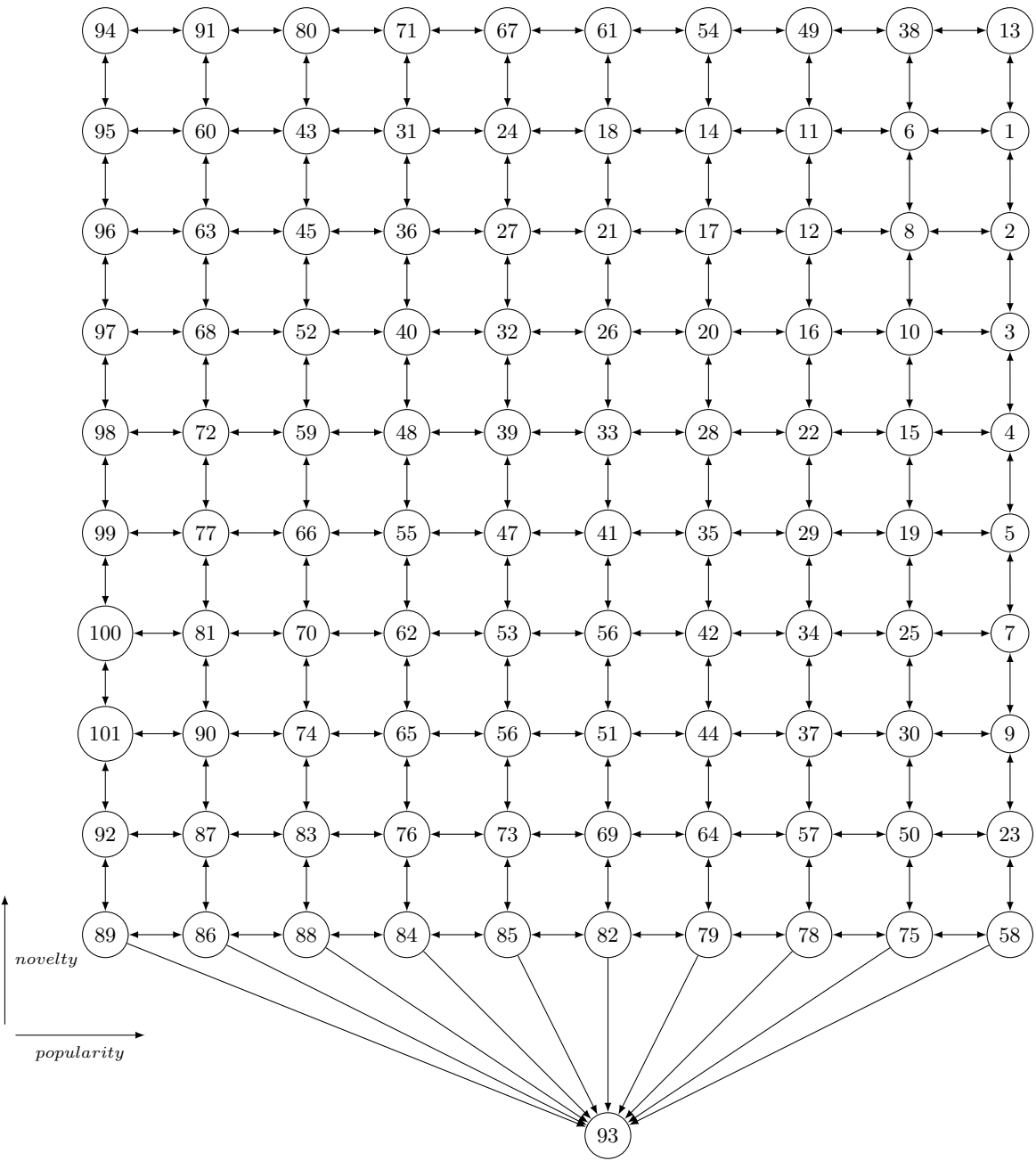

%% file: Evaluation.tex
\section{Evaluation}
\label{sec:evaluation}

Now we validate if the states, transition probabilities and, mainly, the ranking of tweets obtained with the \emph{Huberman-Wu} algorithm guarantee that (i) tweets are arranged according to the utility they will have for the users and (ii) the attention they will receive. With this aim, we validate our approach by measuring the degree of similarity between the ranking of tweets according to their expected utility in $t+1$ (Section~\ref{sec:meth-solution}) and that according to their actual utility (reward) in $t+1$. To perform this evaluation, we consider all the tweets and retweets produced during the last month of observation. In the rest of this section, we detail the experiments conducted and the results obtained.

\subsection{Experiments}

Assuming that the system (Twitter) could update users' timelines at discrete time intervals of one minute and that the \emph{active} tweets in each instant (minute) $t$ --those that could be displayed- are those that have been posted in the last hour, the steps of our validation (in each instant $t$) are the following:

\begin{enumerate}
    \item Find out the state of each tweet in $t$ according to its novelty and popularity and rank the tweets according to their \emph{expected utility} in $t+1$ (ranking of their states obtained in Section~\ref{sec:meth-solution}). 
    \item Find out the state of each tweet in $t+1$ and rank them according to their \emph{actual utility} in $t+1$.
    \item Measure the similarity between the rankings in (1) and (2). 
\end{enumerate}

\vspace{5pt}
\subsection{Results}

In order to measure the level of similarity between the rankings in (1) and (2), we used the well known \emph{Normalized Discounted Cumulative Gain} ($nDCG$)~\cite{jarvelin2002cumulated}, considering the \emph{actual utility} of the tweets in $t+1$ as an indicator of their relevance. The formula of $nDCG$ is given in Equation~\ref{eq:ndcg}:

\begin{equation}
    nDCG = \frac{1}{Z} \displaystyle\sum_{p=1~}^{k} \frac{2^{s(p)}-1}{log(1+p)}
   \label{eq:ndcg} 
\end{equation}

\noindent where $k$ is the number of \emph{active items} in $t$, $s(p)$ is the function that represents the reward 
(actual utility) in $t+1$ given to the tweet at position $p$, and $Z$ is a normalization factor derived from the perfect ranking of tweets that yields a maximum $nDCG$ value of $1$.

\begin{figure*}[t!]
	\begin{tabular}{cc}
    \begin{subfigure}[b]{0.5\textwidth} 
        \centering
        \includegraphics[scale=.45, angle=0]{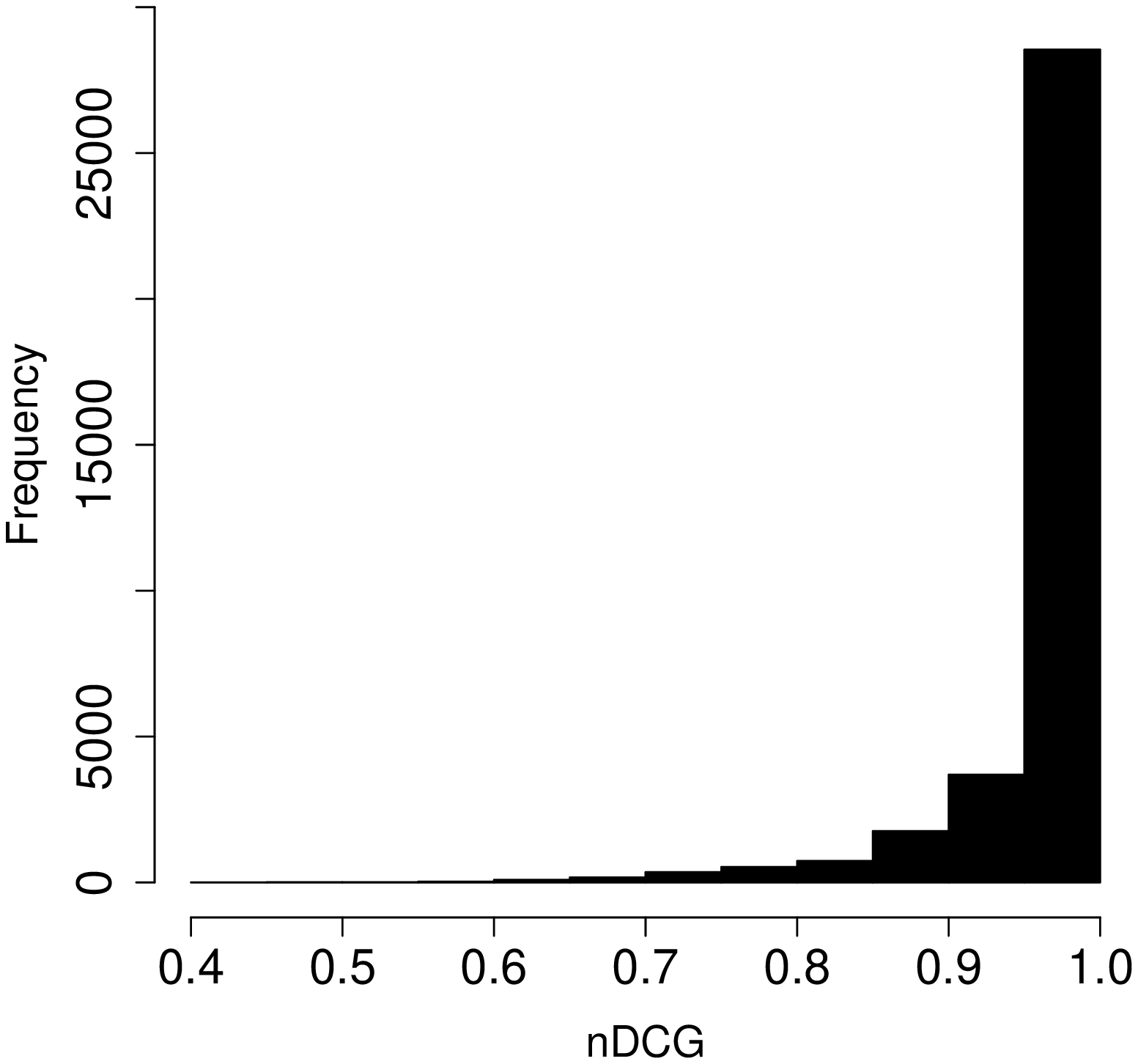}
        \caption{\emph{nDCG} values obtained during each instant of decision.}
        \label{fig:results_a}
    \end{subfigure} &
    \begin{subfigure}[b]{0.5\textwidth}
        \centering
        \includegraphics[scale=.45, angle=0]{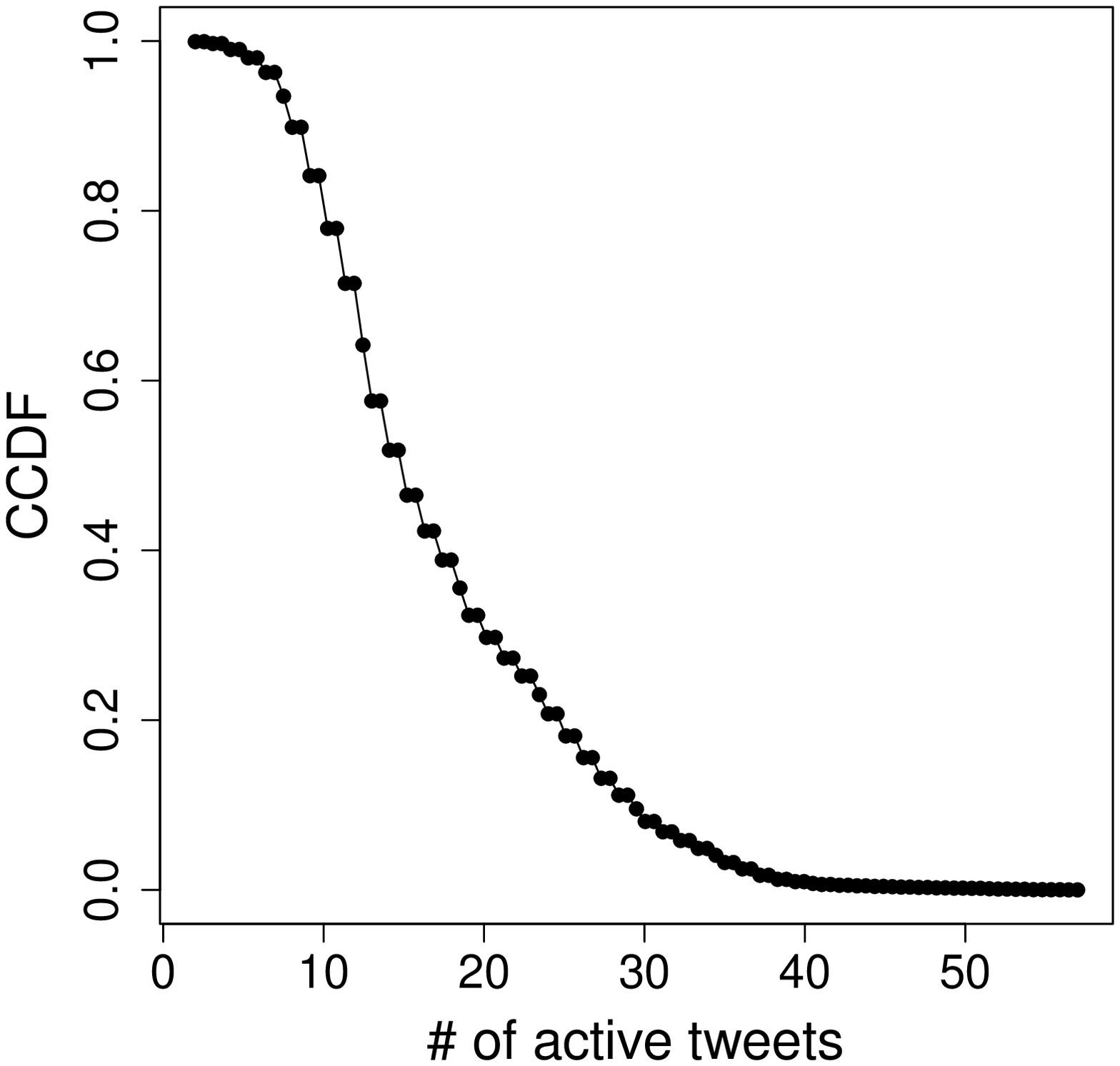}
        \caption{CCDF of the \# of active tweets during each instant of decision.}
        \label{fig:results_b}
    \end{subfigure} \\ 
    \end{tabular}
    \caption{Results.}
      \label{fig:results}
\end{figure*}

Using the utility in $t+1$ as the ground truth for computing the $nDCG$ for each time $t$ ($nDCG=1$ means perfect ranking), the $nDCG$ evaluation result for each instant $t$ is shown in Figure~\ref{fig:results}. Specifically, Figure~\ref{fig:results_a} contains the distribution of the different values of $nDCG$ over time, whereas Figure~\ref{fig:results_b} shows the CCDF of the number of active tweets per minute. The average of $nDCG$ obtained is close to $1$ ($0.97$), which means that our method predicts with high accuracy the expected utility of the tweets. On the other hand, the (Pearson) correlation between the $nDCG$ and the number of \emph{active} tweets in each instant is almost zero ($-0.06$), which shows the independence of the predictive power of our methodology with the number of items (tweets) to rank.

\subsection{Validation considering users' attention}

Now, we check if the ranking obtained with our methodology is such that, in each time $t$, tweets are ranked according to the attention that they will receive in the next time interval $[t,t+1]$. To this aim, and given that the number of users that read every tweet is not available, we approximate the attention of every tweet by the number of retweets, replies and favourites that they have received in the interval $[t,t+1]$. Although this is not accurate, it gives us an approximation of the number of views received by a tweet, since the more people read it, the more prone it is to receive retweets (replies or favourites).

Table~\ref{tab:attention} contains the average and standard deviation of the $nDCG$ for all the instants of decision, considering the attention received by the tweets (number of retweets, replies and favourites) in $[t,t+1]$ as indicators of their relevance. The average $nDCG$ obtained was $0.76$ in the case of only considering the number of retweets received (\# RT in the figure) as indicator of attention, with a standard deviation of $0.21$. Similar values are obtained when the attention is approximated by the sum of retweets and replies received. However, the results are worse in the case of considering also the number of favourites (lower average and higher standard deviation). This is because the temporal distribution of the number of favourites in our dataset is not accurate since we can only retrieve, from the Twitter API, tweets, retweets and replies, but the number of favourites of a tweet must be inferred from the aforementioned entities. 

\begin{table}
    \centering
    \caption{$nDCG$ values considering attention}
    \label{tab:attention}
    \begin{tabular}{|c|c|c|} \hline
            \multicolumn{1}{|c|}{ \scriptsize{}} & \multicolumn{1}{|c|}{ \textbf{\scriptsize{average}}} &
            \multicolumn{1}{|c|}{ \textbf{\scriptsize{std. dev.}}} \\ \hline 
            \# RT & $0.76$ & $0.21$  \\ \hline
            \# RT + \# replies & $0.76$ & $0.21$ \\ \hline
            \# RT + \# replies + \# favourites & $0.69$ & $0.25$  \\ \hline 
        \end{tabular}
  \end{table}

\subsection{Comparison between our ranking and those according to novelty and popularity}

Finally, we compare the arrangement of tweets obtained with our methodology with those according to (i) novelty and (ii) popularity (expressed as number of retweets received) of the tweets in time $t$. To this aim, we compute the $nDCG$ considering the utility and the attention received in the next time interval as indicators of their relevance. Results are displayed in Table~\ref{tab:comparison}. 

\begin{table*}
    \centering
    \caption{Comparison between our arrangement of tweets and the arrangement according to (i) novelty and (ii) popularity of tweets.}
    \label{tab:comparison}
    \noindent\begin{tabularx}{410pt}{|c|X|X|X|}  \hline 
        \multicolumn{1}{|c|}{ \hspace{4.8cm} \scriptsize{}} & \multicolumn{1}{c|} {\centering \hspace{0.78cm} \textbf{\scriptsize{Our method}} \hspace{0.8cm}} &
        \multicolumn{1}{c|}{ \hspace{1.45cm} \textbf{\scriptsize{novelty}} \hspace{1.40cm}} &
        \multicolumn{1}{c|}{ \hspace{1.00cm} \textbf{\scriptsize{popularity}} \hspace{1.00cm}} \\ \hline
     \end{tabularx}\offinterlineskip
      
    \begin{tabularx}{410pt}{|c|c|c|c|c|c|c|} \hline
            
            \multicolumn{1}{|c|}{ \scriptsize{}} & \multicolumn{1}{c|}{ \textbf{\scriptsize{average}}} &
            \multicolumn{1}{c|}{ \textbf{\scriptsize{std. dev.}}} & \multicolumn{1}{c|}{ \textbf{\scriptsize{average}}} &
            \multicolumn{1}{c|}{ \textbf{\scriptsize{std. dev.}}} & \multicolumn{1}{c|}{ \textbf{\scriptsize{average}}} &
            \multicolumn{1}{c|}{ \textbf{\scriptsize{std. dev.}}}\\ \hline 
            Utility & $0.97$ & $0.05$ & $0.85$ & $0.09$ & $0.68$ & $0.11$  \\ \hline 
            \# RT & $0.76$ & $0.21$ & $0.66$ & $0.21$ & $0.46$ & $0.21$ \\ \hline
            \# RT + \# replies & $0.76$ & $0.21$ & $0.67$ & $0.21$ & $0.45$ & $0.21$ \\ \hline
            \# RT + \# replies + \# favourites & $0.69$ & $0.25$ & $0.62$ & $0.24$ & $0.40$ & $0.21$ \\ \hline 
        \end{tabularx}
  \end{table*}

The results show that the ranking of tweets according to the expected utility in $t+1$ obtained with our methodology presents a higher level of similarity with that according to the real utility in $t+1$ (ground truth) than the ranking according to the novelty and, specially, the one according to the popularity of the tweets in $t$. Also, and what it is more important, our ranking of tweets is more similar to the attention received in the interval $[t,t+1]$ than the rankings according to novelty and to popularity. As a conclusion, top-tweets obtained by our methodology are, in average, tweets with higher utility for the users in $t+1$ and will receive more attention in the interval $[t,t+1]$ than the top-tweets in the rankings according to (i) novelty and (ii) popularity in $t$.

As justification for the slight improvement achieved by the arrangement of tweets proposed by our model with respect to the one based on novelty, consider that, in the case of novelty, users see the tweets in such a way that the most recent are ``the most exposed'' (in the top positions) to the users and therefore are more likely to be seen (and therefore retweeted) in the next time interval.

%% file: PeakHours.tex
\section{Analysis considering only peak hours}
\label{sec:peakHours}

We repeated the experiment but considering only those tweets and retweets posted during daily hours in the US -\emph{peak hours}-, hours in which the majority of the tweets in our dataset were posted (see Section~\ref{sec:meth-data}). We fixed the states and the transition probabilities taking into account only those tweets posted between $12am$ and $2am$ of the first month of observation, and validated our proposal with those tweets posted between $12am$ and $2am$ of the second month. We kept the limits between the states according to their novelty ($lim_n = \{ 1, 2, 3, 4, 5, 6, 7, 8, 9, 20, 60 \}$), but the limits according to their popularity were recalculated applying the methodology explained in Section~\ref{sec:st-rew} over the subset of tweets ($lim_p = \{ 0,1,18,25,31,38,47,59,80,133, \infty \}$). This also resulted in new rewards for the states: $r_n = \{ 0.28,1,0.90,0.76, 0.61,0.49,0.41,0.35,0.19,0.06 \}$ and $r_p = \{ 0.01,0.05,0.09,0.13,0.17,0.20,0.24,0.31,0.39,1 \}$, in which the difference between the reward of the \emph{top} state and the other states has been increased with respect to that in the initial analysis (Section~\ref{sec:st-rew}).

Results of this analysis displayed in Table~\ref{tab:comparisonPEAK} show that the average value of the $nDCG$ for the expected utility when considering only peak hours is slightly lower ($0.96$) than when considering all the tweets posted during the whole day ($0.97$). The reason could be that the more tweets used to define the states and calculate the transition probabilities, the more accurate the parameters are and therefore better accuracy in the prediction. On the contrary, the average $nDCG$ for the attention received when considering only peak hours is slightly better than when considering the tweets posted during the whole day ($0.77$ in the case of peak hours in comparison with the $0.76$ when considering all the tweets). Similar results were obtained when measuring the $nDCG$ considering the rankings according to novelty and according to popularity during peak hours, with the ranking according to popularity being the one that achieves the highest improvement when we consider all the tweets. Finally, it is noticeable that there is an increase in the average value of $nDCG$ when tweets posted during peak hours are ranked according to their popularity as compared to its value when all the tweets are ranked.

\begin{table*}
    \centering
    \caption{Comparison between our arrangement of tweets and the arrangement according to (i) novelty and (ii) popularity of tweets (peak hours).}
    \label{tab:comparisonPEAK}
    \noindent\begin{tabularx}{410pt}{|c|X|X|X|}  \hline   
        \multicolumn{1}{|c|}{ \hspace{4.8cm} \scriptsize{}} & \multicolumn{1}{c|} {\centering \hspace{0.78cm} \textbf{\scriptsize{Our method}} \hspace{0.8cm}} &
        \multicolumn{1}{c|}{ \hspace{1.45cm} \textbf{\scriptsize{novelty}} \hspace{1.40cm}} &
        \multicolumn{1}{c|}{ \hspace{1.00cm} \textbf{\scriptsize{popularity}} \hspace{1.00cm}} \\ \hline
     \end{tabularx}\offinterlineskip
      
    \begin{tabularx}{410pt}{|c|c|c|c|c|c|c|} \hline
            
            \multicolumn{1}{|c|}{ \scriptsize{}} & \multicolumn{1}{c|}{ \textbf{\scriptsize{average}}} &
            \multicolumn{1}{c|}{ \textbf{\scriptsize{std. dev.}}} & \multicolumn{1}{c|}{ \textbf{\scriptsize{average}}} &
            \multicolumn{1}{c|}{ \textbf{\scriptsize{std. dev.}}} & \multicolumn{1}{c|}{ \textbf{\scriptsize{average}}} &
            \multicolumn{1}{c|}{ \textbf{\scriptsize{std. dev.}}}\\ \hline 
            Utility & $0.96$ & $0.06$ & $0.82$ & $0.10$ & $0.83$ & $0.11$  \\ \hline 
            \# RT & $0.77$ & $0.22$ & $0.68$ & $0.22$ & $0.58$ & $0.23$ \\ \hline
            \# RT + \# replies & $0.77$ & $0.23$ & $0.67$ & $0.22$ & $0.58$ & $0.23$ \\ \hline
            \# RT + \# replies + \# favourites & $0.72$ & $0.25$ & $0.64$ & $0.24$ & $0.54$ & $0.24$ \\ \hline 
        \end{tabularx}
  \end{table*}

%% file: Related.tex
\section{Related Work}
\label{sec:related}

Since the emergence of the Web, the exponential growth of information has lead to the development of algorithms to rank the information that users receive when make a particular query. Ranking~\cite{baeza1999modern}, therefore, is at the core of the information retrieval (IR) scientific discipline behind search engines as \emph{Google} and \emph{Yahoo! Search}. Although ranking has been studied for decades, a recent trend deals with applying machine learning techniques to learn to rank functions automatically: \emph{Learning to rank}~\cite{liu2009learning}. However, and independently of the algorithm used to rank in IR, the problem we aim to solve is slightly different to the typical problem in IR: on Twitter, like on other social media platforms, tweets are displayed in users' timelines without them have to specify any explicit query in order to obtain relevant tweets from the system.

Closer to our scenario it is the problem of ranking social streams. Users' timelines on social media platforms as \emph{Facebook}, \emph{Twitter} or \emph{LinkedIn} have recently been the natural field of application of algorithms that rank or prioritize content according to the relevance for their owners~\cite{kincaid2010edgerank,duan2010empirical,chen2010short,hong2012learning}. As an outstanding example, the well known Facebook \emph{EdgeRank}~\cite{kincaid2010edgerank} ranks the items in user's \emph{news feed} according to a function that depends on the novelty of the item, the type of item and the strength of the relationship between the user and the creator of the item. Independently of the ranking algorithm, all of these scenarios have in common that (i) items are sorted according to their relevance to the user instead of their relevance to an explicit query as in the traditional IR and (ii) the ranking of the items depends on the owner of the timeline (user-dependent ranking). In order to select or \emph{recommend} the most attractive and relevant content to the user, these solutions often use personalization techniques based mainly on matching users' profiles (previously obtained) with items' content~\cite{diaz2012real,uysal2011user,feng2013retweet,abel2011analyzing,de2012chatter}. Contrarily, our proposal aims to provide a ranking of tweets that maximise the utility or relevance for \emph{all} Twitter users, and not only for a specific user or group of users. 

Moreover, although in our model we consider retweets as both indicators of popularity and attention, the aim of our study is not to predict if a given user will retweet a tweet or how many times a tweet will be retweeted~\cite{boyd2010tweet,yang2010understanding,peng2011retweet}, but to automatically decide, on the basis of its novelty and popularity, if a tweet should be shown in a top or a bottom position of users' timelines.

%% file: Conclusion.tex
\section{Conclusion}
\label{sec:conclusion}

Social media platforms, and in particular the microbologging service Twitter, are examples of competing attention environments whose subscribers have to decide what content to prioritize to their followers in order to get the most attention. The fact that Twitter displays tweets in decreasing order of publication limits the capability of tweet promotion, since the more recent tweets are the ones to be displayed in the top positions and the time they remain in these positions only depends on the number of tweets posted in the following minutes. Therefore, tweets that users' find valuable are frequently hidden. As a solution, we customized the \emph{Huberman-Wu} algorithm~\cite{huberman2008economics} to select the optimal arrangement of tweets that maximises utility for the users. We experimentally validated this method using real data from Twitter. Our results confirmed the suitability of using the aforementioned algorithm in order to, not only arrange the tweets according to their utility for the users, but also maximize users' attention.

So far, we have focused our study on Twitter, but the extension of this mechanism to other social news aggregators, such as \emph{\url{digg.com}} or \emph{\url{reddit.com}} is straightforward. Also, although we only consider a particular kind of tweets (those that contain news), this methodology could also be extended to other kinds of tweets such as promotional campaigns and status updates from users. This would require dealing with different temporal patterns of retweet behaviour. And, the more patterns exist, the more different systems must be considered (in terms of definition of states and calculation of their transition probabilities).

Finally, in our model we have taken into account the retweets made by all Twitter users independent of the specific interests of the users we display the items to. As future work, we plan to add personalization to our methodology to select the tweets to display. That is, different arrangement of tweets will be displayed to different users --groups of users-- on the basis of their interests, the interests of their friends or the interests of users with similar interests (users that historically have retweeted --replied, marked as favourites-- the same tweets) as the given user.